\documentclass[aps,prl,twocolumn,superscriptaddress,showpacs,floatfix]{revtex4}

\usepackage{graphicx}
\usepackage{epstopdf}
\usepackage{dcolumn}
\usepackage{bm}
\usepackage{float}
\usepackage{braket}

\begin{document}

\title {Giant atomic displacement induced by built-in strain in metastable Mn$_3$O$_4$}

\author{S. Hirai}
\affiliation{Department of Geological and Environmental Sciences, Stanford University, California 94305, USA}
\affiliation{Stanford Institute of Energy and Materials Science, SLAC National Accelerator Laboratory, 2575 Sand Hill Road, Menlo Park, CA 94025, USA}

\author{A. M. dos Santos}
\affiliation{Oak Ridge National Laboratory, 1 Bethel Valley Road, Oak Ridge, Tennessee 37831, USA}

\author{M. C. Shapiro}
\affiliation{Stanford Institute of Energy and Materials Science, SLAC National Accelerator Laboratory, 2575 Sand Hill Road, Menlo Park, CA 94025, USA}
\affiliation{Geballe Laboratory for Advanced Materials and Department of Applied Physics, Stanford University, Stanford, CA 94305, USA}

\author{J. J. Molaison}
\affiliation{Oak Ridge National Laboratory, 1 Bethel Valley Road, Oak Ridge, Tennessee 37831, USA}

\author{N. Pradhan}
\affiliation{Oak Ridge National Laboratory, 1 Bethel Valley Road, Oak Ridge, Tennessee 37831, USA}

\author{M. Guthrie}
\affiliation{Oak Ridge National Laboratory, 1 Bethel Valley Road, Oak Ridge, Tennessee 37831, USA}
\affiliation{Carnegie Institution of Washington, 5251 Broadbranch Road, N.W., Washington, D.C. 20015, USA}

\author{C. A. Tulk}
\affiliation{Oak Ridge National Laboratory, 1 Bethel Valley Road, Oak Ridge, Tennessee 37831, USA}

\author{I. R. Fisher}
\affiliation{Stanford Institute of Energy and Materials Science, SLAC National Accelerator Laboratory, 2575 Sand Hill Road, Menlo Park, CA 94025, USA}
\affiliation{Geballe Laboratory for Advanced Materials and Department of Applied Physics, Stanford University, Stanford, CA 94305, USA}

\author{W. L. Mao}
\affiliation{Department of Geological and Environmental Sciences, Stanford University, California 94305, USA}
\affiliation{Stanford Institute of Energy and Materials Science, SLAC National Accelerator Laboratory, 2575 Sand Hill Road, Menlo Park, CA 94025, USA}

\begin{abstract}
We present x-ray, neutron scattering and heat capacity data that reveal a coupled first-order magnetic and structural phase transition of the metastable mixed-valence post-spinel compound Mn$_3$O$_4$ at 210 K. Powder neutron diffraction measurements reveal a magnetic structure in which Mn$^{3+}$ spins align antiferromagnetically along the edge-sharing \emph{a}-axis, with a magnetic propagation vector k = [1/2, 0, 0]. In contrast, the Mn$^{2+}$ spins, which are geometrically frustrated, do not order until a much lower temperature. Although the Mn$^{2+}$ spins do not directly participate in the magnetic phase transition at 210 K, structural refinements reveal a large atomic shift at this phase transition, corresponding to a physical motion of approximately 0.25 {\AA} even though the crystal symmetry remains unchanged. This "giant" response is due to the coupled effect of built-in strain in the metastable post-spinel structure with the orbital realignment of the Mn$^{3+}$ ion. 

\end{abstract}


\maketitle

\narrowtext

Large atomic movements can occur at displacive phase transitions. For example, atomic displacements of 0.05 - 0.4 {\AA} are found for many ferroelectric materials, which lose their inversion symmetry \cite{wyck86}. Typically, atomic displacements are much smaller for isostructural transitions, although it has recently been shown that a "giant" atomic displacement of 0.05 - 0.09 {\AA} occurs in hexagonal manganites due to large magnetoelastic coupling \cite{slee08}. In the present work, we obtain a much larger atomic displacement up to 0.25 {\AA} at an isostructural coupled magnetic and structural phase transition in metastable Mn$_3$O$_4$ post-spinel, which is distinct from the well known magnetostriction. This effect is a direct consequence of the highly strained metastable crystal lattice of Mn$_3$O$_4$ in the post-spinel structure, and provides a new avenue for the design of materials exhibiting giant atomic displacements without breaking inversion symmetry.

\begin{figure}[!h]
\centerline{\includegraphics[scale=1]{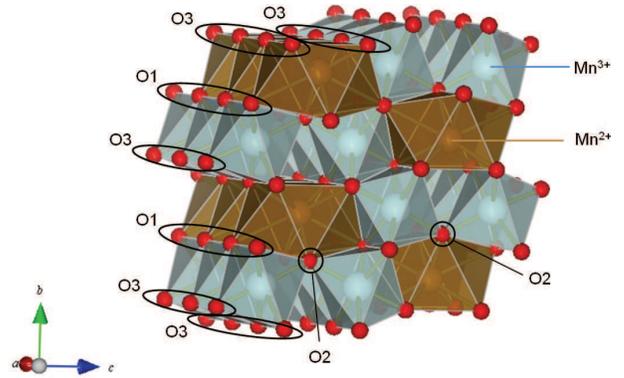}}
\caption{\label{fig:mn3o41}
(color online) Crystal structure of metastable Mn$_3$O$_4$ post-spinel (orthorhombic, \emph{Pbcm}(57)) at ambient pressure and room temperature. Gold shading represents the Mn$^{2+}$O$_8$ coordination polyhedra, while blue shading represents the Mn$^{3+}$O$_6$ octahedra. Large atomic displacements at the magnetic phase transition are a consequence of the the built-in strain in this metastable structure.}
\end{figure}

Mn$_3$O$_4$ is a unique mixed-valence oxide which adopts a tetragonally distorted spinel structure at ambient conditions \cite{amin26}. This spinel phase undergoes a structural phase transition at 15 GPa into the CaMn$_2$O$_4$-type phase, referred to as post-spinel (Figure \ref{fig:mn3o41}), which is quenchable to ambient pressure \cite{paris92, merlin10}. The magnetic properties of Mn$_3$O$_4$ spinel have been extensively studied, with three magnetic transitions and a pronounced magnetodielectric coupling being reported \cite{suzuki08, tack07}. On the other hand, studies of Mn$_3$O$_4$ post-spinel are limited to structural work using X-ray diffraction (XRD) \cite{paris92}. In the present work, we have determined the magnetic structure of Mn$_3$O$_4$ post-spinel at ambient pressure using neutron diffraction. We find that the onset of long range magnetic order of the Mn$^{3+}$ moments at $T_N = 210$ K is coupled to a first order structural transition. The Mn$^{2+}$ moments do not order at $T_N$ due to the intrinsic frustration of the crystal lattice, but nevertheless experience a large atomic shift of 0.25 {\AA}. This "giant" effect is due to the exaggerated strain of the Mn$^{2+}$O$_8$ polyhedra in the metastable post-spinel structure. 

\begin{figure}[!h]
\centerline{\includegraphics[scale=1]{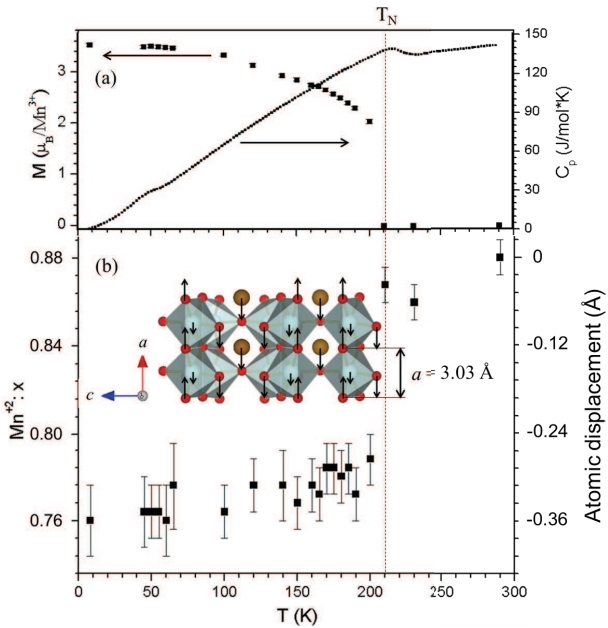}}
\caption{\label{fig:mn3o43}
(color online) Temperature dependence of (a) the magnetic moment of Mn$^{3+}$ (left axis), the specific heat $C_p$ (right axis); and (b) the atomic position of Mn$^{2+}$ in Mn$_3$O$_4$ post-spinel, expressed on the left axis in terms of the atomic coordinate x in the unit cell, and on the right axis in terms of the actual atomic displacement relative to the position at 290 K. Dashed vertical line indicates approximate transition temperature. Inset shows the atomic displacements at 200 K found upon cooling below $T_N = 210$ K for Mn$^{2+}$ (8 \% shift in unit cell coordinates, corresponding to 0.25 {\AA}) and O3 (5 \% shift in unit cell coordinates, corresponding to 0.16 {\AA}) represented by black arrows. The much smaller atomic displacement of Mn$^{3+}$ (2 \% shift in unit cell coordinates, corresponding to 0.06 {\AA}) due to the magnetoelastic coupling is also shown. The directons of atomic displacement shown in the figure are only the representaive ones, and equal number of atoms move in the opposite directions due to the \emph{Pbcm} symmetry. The arrows are illustrated in proportional length to the atomic displacements of Mn$^{2+}$, O3 and Mn$^{3+}$, but are all exaggerated by a factor of 4 for clarity.}
\end{figure}

Polycrystalline samples of Mn$_3$O$_4$ post-spinel were synthesized using a Paris-Edinburgh cell apparatus. Mn$_3$O$_4$ spinel was used as a starting material, which was originally prepared by heating MnCO$_3$ in air at 1400 K for 16 hours. The starting material was pressurized up to 20 GPa using a cell fitted with polycrystalline diamond double toroidal anvils \cite{klotz95}, followed by a slow decompression at a rate of 5 GPa/hour to ambient pressure. The target pressure was set well above the transition pressure (15 GPa) in order to ensure full transformation of the starting material into the post-spinel phase. Low temperature and room temperature neutron diffraction measurements were conducted at beamline 3 (SNAP) of the Spallation Neutron Source (SNS), Oak Ridge National Laboratory (ORNL) on the recovered sample at elevated temperatures between 290 K and 8 K in the \emph{d}-spacing range of 0.5 - 6 {\AA}.  Since the atomic positions of octahedrally coordinated Mn$^{3+}$ (d$^4$: $t_2$$_g$$^3$ $e_g$$^1$) and eight-fold coordinated Mn$^{2+}$ (d$^5$: $t_2$$_g$$^3$ $e_g$$^2$) ions are crucial for determining the magnetic structure, we also confirmed the lattice parameters obtained from neutron diffraction using synchrotron XRD. Room temperature XRD data were collected at beamline 16-BMD of the Advanced Photon Source (APS) in Argonne National Laboratory (ANL)using x-rays with a wavelength of $\lambda = 0.41222$ {\AA}. GSAS software was used for Rietveld refinement of the crystal and magnetic structures (Figure \ref{fig:mn3o41}, Figure \ref{fig:mn3o43}). Lattice constants obtained from the synchrotron XRD measurements were used as a starting model for the neutron diffraction refinement, while the atomic positions obtained from the neutron diffraction were used as a starting model for the XRD refinement, because neutron diffraction has better powder averaging (due to larger beam size) and more sensitive to the oxygen positions. In the following discussion we first describe the room temperature refinements, and then the low temperature results. 

\begin{figure}[!h]
\centerline{\includegraphics[scale=1.1]{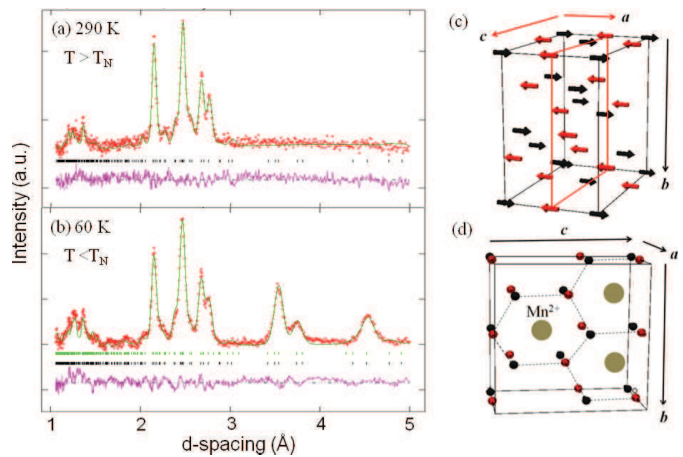}}
\caption{\label{fig:mn3o42}
(color online) Neutron powder diffraction profile of Mn$_3$O$_4$ post-spinel at (a) 290 K and (b) 60 K. Magnetic Bragg peaks are indicated by green marks, while nuclear Bragg peaks are indicated by black marks. The pink curve represents the difference between the observed and calculated neutron diffraction intensities based upon the refined magnetic structure described in the main text. (c) Magnetic structure for Mn$^{3+}$ spins in Mn$_3$O$_4$ (arrows represent Mn$^{3+}$ spins, black arrows are the spins parallel to the spin at the origin, while red arrows are the spins antiparallel to the spin at the origin). (d) Magnetic frustration of the Mn$^{2+}$ spins against the antiferromagnetically ordered Mn$^{3+}$ spins (black signs point out of the paper, while red signs point into the paper). The gold sphere represents the Mn$^{2+}$ ion which is situated at the center of the honeycomb composed of six antiferromagnetically ordered Mn$^{3+}$ spins.}
\end{figure}

Powder neutron diffraction and XRD refinement at room temperature (Figure \ref{fig:mn3o42}(a)) demonstrate that the quenched sample of Mn$_3$O$_4$ crystallizes in the CaMn$_2$O$_4$-type post-spinel structure with \emph{Pbcm} space group as anticipated following previous reports \cite{paris92}. The lattice parameters obtained by these two methods were consistent, the difference being within the standard deviation. The room-temperature crystal structure, in which Mn$^{3+}$O$_6$ octahedra form one-dimensional edge-sharing chains along the \emph{a}-axis, and zigzag along the \emph{b}-axis, is shown in Figure \ref{fig:mn3o41}. Eight-fold coordinated Mn$^{2+}$ (d$^5$: $t_2$$_g$$^3$ $e_g$$^2$) ions are situated in the cavity provided by the zigzag connectivity of the Mn$^{3+}$O$_6$ octahedra. Detailed crystallographic data of atomic coordinates and bond lengths/angles are shown in Table \ref{tab:paramtab}. In particular, we note that at room temperature, the Mn$^{3+}$O$_6$ octahedron comprises two short, two medium and two long Mn$^{3+}$-O bond distances due to the anisotropic connectivity of the coordination-octahedra.

\begin{table*}[tp]
\centering
\begin{tabular}{| c | c | c|c  | c  | c  | l        | l        |l        |}
  \hline
                T (K)   & \emph{a} ({\AA})  & \emph{b} ({\AA})&\emph{c} ({\AA})  &  Mn$^{2+}$: x  & O3: x   & Mn$^{3+}$-O(m) ({\AA})        & Mn$^{3+}$-O(s) ({\AA})      & Mn$^{3+}$-O(l) ({\AA})    \\
  \hline
                 290   &  3.032(1)  & 9.880(4) & 9.583(3)        & 0.876(4)   & 0.724(2)   & 1.996(12), 2.050(17) & 1.860(14), 1.909(5) & 2.223(18), 2.250(14) \\
                  60   &  3.020(2)  & 9.842(6) & 9.568(4)        & 0.760(8)   & 0.800(2)   & 2.096(11), 2.097(18) & 1.885(13), 1.899(7) & 2.127(11), 2.157(18) \\

  \hline
\end{tabular}
\caption{Table of crystallographic information of Mn$_3$O$_4$ post-spinel for representative temperatures above (290 K) and below (60 K) $T_N = 210$ K. The middle, short and long Mn$^{3+}$-O bond distances are abbreviated as m, s and l, respectively.}
\label{tab:paramtab}
\end{table*}

Powder neutron diffraction measurements at low temperature reveal a coupled magnetic and structural phase transition at $T_N = 210$ K. Magnetic peaks were found by excluding the nuclear peaks ($<3$ {\AA}), and the propagation vector of k = [1/2, 0, 0] for the magnetic unit cell was found to have the best fit using supercell in Fullprof software (representative data shown in Figure \ref{fig:mn3o42}(b) for a temperature of 60 K). In other words, below 210 K, Mn$_3$O$_4$ post-spinel adopts an \emph{a}-axis doubled magnetic unit cell (Figure \ref{fig:mn3o42}(c)), in which Mn$^{3+}$ spins are aligned antiferromagnetically along the edge sharing \emph{a}-axis. Refinement was performed based on this magnetic unit cell in the \emph{Pbca} space group, yielding an ordered moment for the Mn$^{3+}$ spins of $\mu_x$ = 3.49(2) $\mu_B$ at 60 K (moment along the crystallographic \emph{a}-axis). The temperature dependence of the ordered moment, shown in Figure \ref{fig:mn3o43}(a), exhibits a rapid drop at $T_N = 210$ K, indicative of a first order phase transition. In contrast, as described in greater detail below, the Mn$^{2+}$ moments do not appear to order at $T_N$, but remain paramagnetic down to approximately 50 K. Based on the refinement results described above, the temperature-dependence of the lattice parameters was also determined for a cooling cycle down to 8 K. For the lattice constants \emph{a} and \emph{b}, a discontinuity around 210 K is clearly observed, signalling a coupled first order structural phase transition, which is accompanied by a lattice volume drop of 0.5(1) \%. In spite of the considerable volume drop, this transition does not change the original \emph{Pbcm} symmetry and the material retains its inversion symmetry. 

Heat capacity ($C_p$) measurements were made for similar polycrystalline samples of Mn$_3$O$_4$ post-spinel between 2 and 290 K, and representative data are shown in Figure \ref{fig:mn3o43}(a). These measurements reveal a broad feature near 210 K, supporting the x-ray and neutron scattering evidence for a phase transition. The width of the "peak" in the heat capacity at this transition likely reflects the presence of inhomogeneous strain quenched from the high pressure synthesis. A similar broadening of an otherwise sharp first order transition has been observed for polycrystalline samples of other spinel compounds, such as MnV$_2$O$_4$ \cite{zhou07}.

Mn$^{2+}$ ions occupy an inequivalent site to Mn$^{3+}$ in the Mn$_3$O$_4$ crystal lattice. Consequently, any magnetic order associated with the Mn$^{2+}$ ions must either lead to additional diffraction peaks for a $Q\neq0$ structure, or an enhancement of the nuclear peaks for a $Q = 0$ structure. Neither  effect is observed down to 55 K, and we conclude that the Mn$^{2+}$ ions do not order until the lower temperature. This effect can be attributed to geometric frustration since the Mn$^{2+}$ ions are situated at the center of a honeycomb arrangement of Mn$^{3+}$ spins that are antiferromagnetically ordered (illustrated in Figure \ref{fig:mn3o42}(d)). However, at 55 K an additional feature is seen in the heat capacity(Figure \ref{fig:mn3o43}(a)), suggesting an additional phase transition. The crystal lattice does not change appreciably at this temperature, nor does the magnetic structure of the Mn$^{3+}$ moments. However, neutron diffraction data reveal new broad and asymmetric magnetic peaks around 1.8 {\AA}, 4.5 {\AA}, 4.9 {\AA} and 5.1 {\AA}(see supplemental material \cite{hirai12}), consistent with the onset of short-range order of the Mn$^{2+}$ moments \cite{hirai12}. Susceptibility measurements reveal the onset of a pronounced hysteresis below this temperature, indicative of either ferromagnetic or glassy behavior \cite{hirai12}. In the following discussion, we focus on the temperature regime above 55 K, in which the Mn$^{2+}$ moments remain paramagnetic.

\begin{figure}[!h]
\centerline{\includegraphics[scale=1.1]{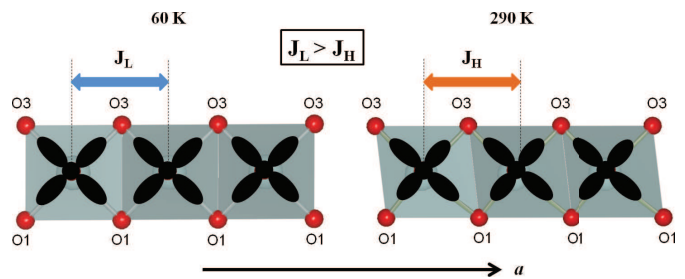}}
\caption{\label{fig:mn3o44}
(color online) Schematic view of the edge-sharing Mn$^{3+}$O$_6$ octahedra along the \emph{a}-axis at 60 K (below $T_N$) and at 290 K (above $T_N$). Blue shading represents the coordination octahedra, and red spheres represent the oxygen ions. The associated $d_x$$_2$$_-$$_y$$_2$ orbitals are shown in black. Above $T_N$, $d_x$$_2$$_-$$_y$$_2$ and d$_3$$_z$$_2$$_-$$_r$$_2$ orbitals are mixed and \emph{d} orbitals partially align along the longest bonds of the Mn$^{3+}$O$_6$ octahedron. The Mn$^{3+}$O$_6$ octahedra are less distorted in the xy plane below $T_N$.  The associated change in the orbital character leads to an enhanced exchange interaction along the \emph{a}-axis.}
\end{figure}

The small changes in lattice parameter and unit cell volume at $T_N$ mask some startlingly large atomic displacements within the unit cell. Atomic coordinates and the associated bond lengths and angles are shown in Table \ref{tab:paramtab} for representative temperatures above (290 K) and below (60 K) $T_N$. More detailed crystallographic parameters are given in the Supplemental Material \cite{hirai12}. Focussing first on the Mn$^{3+}$O$_6$ coordination octahedra, the structural transition involves a relatively large motion of the O3 oxygen ion, with a change of 5 \% in the unit cell coordinate, corresponding to a displacement of 0.16 {\AA}. This motion leads to a smaller distortion of the octahedron in the xy plane relative to the room temperature structure, such that at low temperatures the four coordinating oxygen ions in the xy plane form an almost perfect square(Figure \ref{fig:mn3o44}). The corresponding change in the orbital character of the Mn$^{3+}$ ion affects the magnitude of the exchange interaction between nearest neighbour Mn$^{3+}$ ions, possibly providing the driving force for the O3 ionic motion. 

Surprisingly, our refinements also reveal a very large atomic shift along the \emph{a}-axis for the Mn$^{2+}$ ions (Figure \ref{fig:mn3o43}(b)) at $T_N$. The unit cell coordinate changes by 8 \%, corresponding to a displacement of 0.25 {\AA} , \textit{even though these ions do not directly participate in the magnetic phase transition}. This effect is even larger than that found in giant magneto-elastic compounds such as hexagonal manganites (atomic displacement: 0.05 - 0.09 {\AA}) \cite{slee08}. However, in the case of Mn$_3$O$_4$ the effect does not arise directly from magnetoelastic coupling, because the Mn$^{2+}$ ions do not directly participate in the magnetic phase transition. The origin of this effect is intimately linked to the metastable nature of Mn$_3$O$_4$ in the post-spinel structure at ambient pressure, and in particular to the severely strained Mn$^{2+}$O$_8$ polyhedra. Mn$_3$O$_4$ post-spinel is isostructural with CaMn$_2$O$_4$, which is thermodynamically stable, but the Mn$^{2+}$ ion is much smaller than Ca$^{2+}$. As a result, the Mn$^{2+}$O$_8$ polyhedron is considerably stretched relative to the ideal bond lengths. This is borne out by the  bond valence sum (BVS), which  is 1.82, smaller than the formal valence 2. The associated strain is rather severe, and can act to magnify the effects of more modest ionic displacements at the magnetic phase transition. Specifically, on cooling below $T_N$, the associated change in position of the O3 ions described above, would result in a further reduction of the BVS of the Mn$^{2+}$ ions if they did not move. Indeed, if the Mn$^{2+}$ ions did not move at $T_N$, the BVS would change from 1.83 to 1.73, far below the optimal value of 2 associated with the formal 2+ valence. Consequently, in order to minimize the elastic energy, the Mn$^{2+}$ ions experience a giant atomic displacement, recovering a low temperature BVS of 1.83, closer to the expected value. Such a large atomic shift of Mn$^{2+}$ suggests that Mn$_3$O$_4$ post-spinel is at the edge of the phase boundary between the post-spinel and the spinel phase and is therefore highly susceptible to large atomic displacements. 

Since Mn$_3$O$_4$ post-spinel is isostructural with CaMn$_2$O$_4$, and the only chemical difference is that Ca$^{2+}$ is replaced by Mn$^{2+}$, it is instructive to compare the physical properties of these two compounds. Mn$_3$O$_4$ exhibits an insulating behavior similar to CaMn$_2$O$_4$ \cite{white09} with a very large resisitivity. CaMn$_2$O$_4$ exhibits long-range antiferromagnetic order below $T_N$= 217 K \cite{white09}, comparable to that of Mn$_3$O$_4$, and also adopts the same magnetic structure as the Mn$^{3+}$ spins for Mn$_3$O$_4$. However, in contrast with Mn$_3$O$_4$ post-spinel, the magnetic phase transition in CaMn$_2$O$_4$ is continuous and the associated atomic displacements of Ca$^{2+}$ and O3 ions are considerably smaller(the corresponding differences in atomic positions between 300 K and 20 K are just 0.012 {\AA} for Ca$^{2+}$, and 0.003 {\AA} for O3 \cite{ling01}, relative to 0.25 {\AA} and 0.16 {\AA} for Mn$^{2+}$ and O3 ions in Mn$_3$O$_4$ post-spinel). The origin of this difference lies in the metastable nature of post-spinel Mn$_3$O$_4$. CaMn$_2$O$_4$ is a thermodynamically stable phase, and the specific crystal lattice is a minimum of the free energy. The lattice is relatively stiff, and associated motion of O3 ions in response to the onset of long-range magnetic order is correspondingly small. However, the built-in strain associated with the metastable post-spinel structure adopted by Mn$_3$O$_4$ quenched from high-pressure leads to a softer crystal lattice, and hence an exaggerated response of both the O3 and Mn$^{2+}$ ions to the onset of long-range magnetic order. 

In conclusion, Mn$_3$O$_4$ in the post-spinel structure at ambient pressure undergoes an isostructural coupled magnetic and structural phase transition at $T_N = 210$ K. The built-in strain associated with the metastable structure leads to a "giant" atomic displacement of the Mn$^{2+}$ ions at $T_N$, even though they do not directly participate in the magnetic order due to the effects of geometric frustration. This novel effect illustrates an alternative avenue to achieve giant atomic displacements coupled to magnetic phase transitions. 

This research is funded by the U.S. Department of Energy (DOE), Office of Basic Energy Sciences (BES). S.H., W.L.M., M.S. and I.R.F. are supported by the U.S. Department of Energy (DOE), Office of Basic Energy Sciences (BES), Division of Materials Sciences and Engineering, under Contact No. DE-AC02-76SF00515. M.G. is supported by EFree, an Energy Frontier Research Center funded by DOE-BES. Neutron diffraction experiments at the SNAP facility were supported by SNS and Center of Nanophase Materials Science of ORNL. XRD experiments were performed at HPCAT, APS, ANL. HPCAT is supported by CIW, CDAC, UNLV, and LLNL through funding from DOE-NNSA, DOE-BES, and NSF. APS is supported by DOE-BES, under Contract No. DE-AC02-06CH11357. We thank D. Ikuta and W. Yang for help with the XRD experiments.

\end{document}